\documentclass[twocolumn,prl,amsmath,amssymb]{revtex4-1}
\usepackage[latin9]{inputenc}
\usepackage{amsmath}
\usepackage{amssymb}
\usepackage{graphicx}
\usepackage[unicode=true,pdfusetitle,
 bookmarks=true,bookmarksnumbered=false,bookmarksopen=false,
 breaklinks=false,pdfborder={0 0 1},backref=false,colorlinks=false]
 {hyperref}
\begin{document}
\title{Anomalous thermodynamics of lattice Bose gases in optical cavities }
\author{Liang He}
\email{liang.he@scnu.edu.cn}

\affiliation{Guangdong Provincial Key Laboratory of Quantum Engineering and Quantum
Materials, SPTE, South China Normal University, Guangzhou 510006,
China}
\author{Su Yi}
\email{syi@itp.ac.cn}

\affiliation{CAS Key Laboratory of Theoretical Physics, Institute of Theoretical
Physics, Chinese Academy of Sciences, Beijing 100190, China}
\affiliation{School of Physical Sciences \& CAS Center for Excellence in Topological
Quantum Computation, University of Chinese Academy of Sciences, Beijing
100049, China}
\begin{abstract}
We investigate thermodynamic properties of lattice Bose gases in optical
cavities in the Mott-insulator limit. We find the system assumes anomalous
thermodynamic behavior that can be traced back to the breaking of
fundamental additivity by its infinite-long range interaction. Specifically,
the system shows striking ensemble inequivalence between the canonical
ensemble and the grand canonical one, sharply manifesting in the distinct
anomalous structure of the thermodynamic phase diagram in the canonical
ensemble. In particular, in the temperature regime around half of
the on-site energy, the system manifests negative compressibility
and anomalous reentrant phase transitions where the ordered charge
density wave phase revives from the disordered homogenous phase upon
increasing the temperature. Direct experimental observation of the
anomalous behavior can be realized in the current experiments with
well-controlled total particle number fluctuations. 
\end{abstract}
\maketitle
Systems with long range interactions can give rise to rich phases
of matter and physical phenomena that are absent in their short range
counterparts. Recently, in the field of ultracold atom physics, various
systems with long-range interactions have been realized in experiments,
including, for instance, quantum gases with large magnetic or electric
dipole moments \citep{Stuhler_PRL_2005,Ni_Science_2008}, ultracold
Rydberg atoms \citep{Heidemann_PRL_2008}, and ultracold gases in
cavities with cavity-photon-mediated interactions \citep{Baumann_nature_2010,Mottl_Science_2012,Landig_Nature_2016}.
Interesting phases such as charge and spin density waves, supersolids,
and topological phases \citep{Goral_PRL_2002,Scarola_PRL_2005,Yi_PRL_2007,Dalla_Torre_PRL_2006,He_PRA_2011,Dutta_Rep_Prog_Phys_2015},
have been predicated to exist in these systems and some of them have
been observed in experiments \citep{Baumann_nature_2010,Mottl_Science_2012,Landig_Nature_2016}.

Comparing the range of interactions in these systems, in fact one
immediately notices that ultracold gases in cavities are quite unique
in the sense of their cavity-photon-mediated interactions being \emph{infinite}-long-range
(ILR)\citep{Baumann_nature_2010,Mottl_Science_2012,Landig_Nature_2016}.
On the one hand, this can give rise to rich phases in ultracold gases
in cavities that share common features with other long-range interacting
systems as what has been revealed by recent experimental and theoretical
investigations \citep{Mottl_Science_2012,Baumann_nature_2010,Landig_Nature_2016,Li_PRA_2013,Habibian_PRL_2013,Chen_PRA_2016,Dogra_PRA_2016,Niederle_PRA_2016,Sundar_PRA_2016,Panas_PRB_2017,Liao_PRA_2018}.
However, on the other hand, from a more fundamental perspective, ultracold
gases in optical cavities in fact holds a distinct position among
other long-range interacting ultracold atom systems. This is due to
the fact that for systems whose range of interactions are sufficiently
long, the fundamental additivity, namely, the sum of the energy of
macroscopic subsystems equals to the energy of the whole system, are
broken by the long-range nature of their interactions. It can make
these systems manifest anomalous thermodynamic behavior which are
absent in their relatively-short-range counterparts \citep{Campa_Phys_Rep_2009,Bouchet_Physica_A_2010},
with one typical example being negative specific heat in gravitational
systems \citep{Lynden_Bell_Mon_Not_R_Astr_Soc_1968}. In this regard,
ultracold gases in optical cavities give rise to the novel scenario
for statistical mechanics of non-additive systems realized in concrete
experimental platforms, and pose the fundamental question of the physical
consequences associated to additivity breaking.

\begin{figure}
\includegraphics[width=3in]{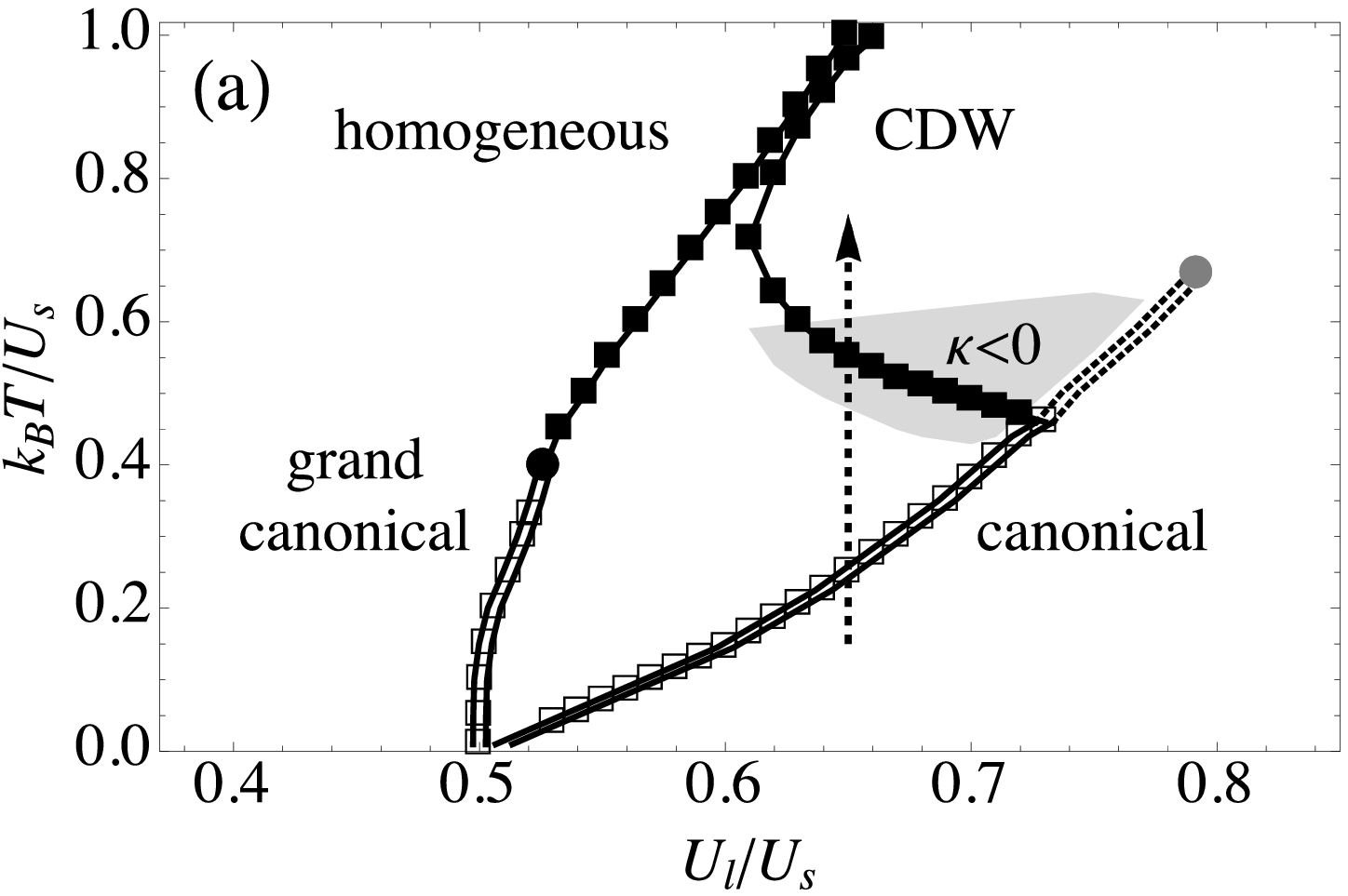}

\includegraphics[width=3.3in]{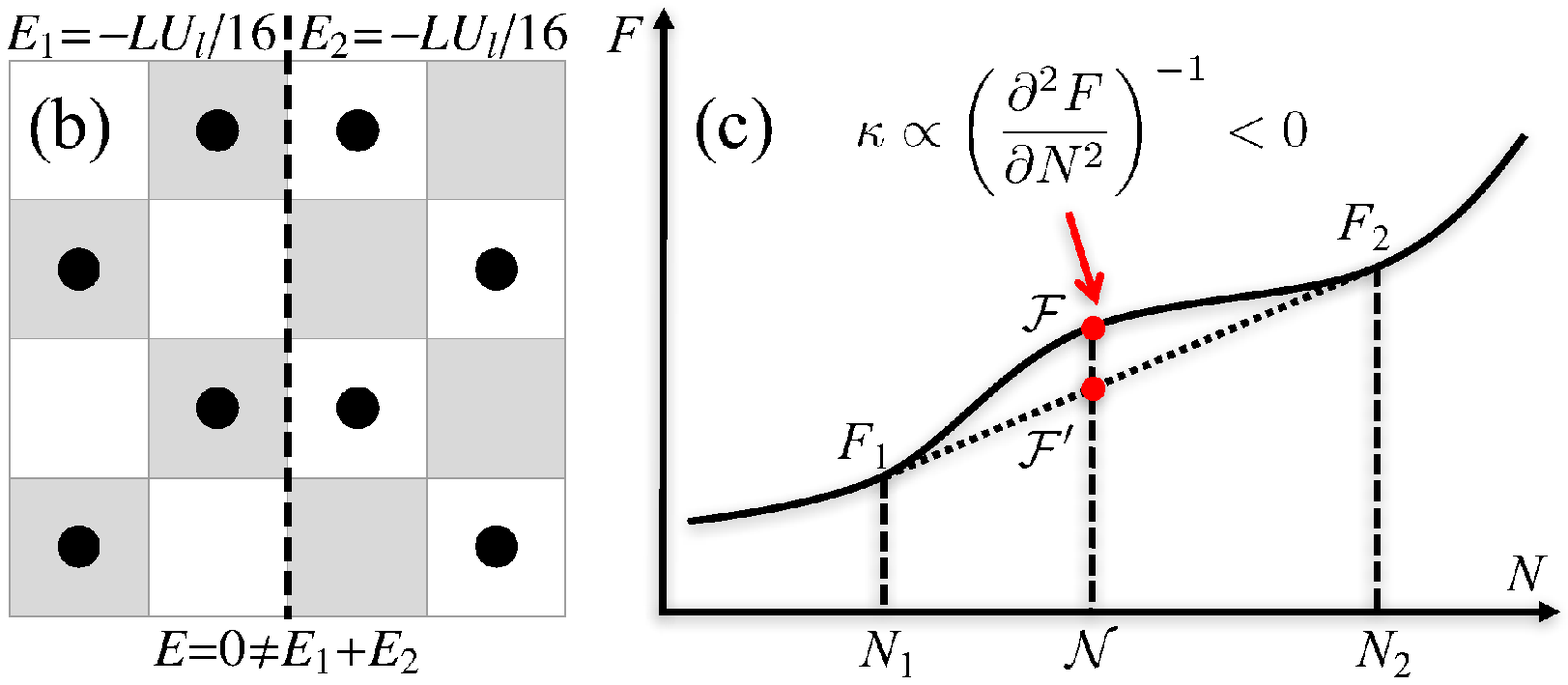}

\caption{(a) Phase diagram of the system at unit filling in the canonical ensemble
showing distinct anomalous structure compared to its GC one. The curve
on the left is the phase boundary in the GC ensemble (reproduced according
to Ref.~\citep{He_arXiv_2020}), which separates the homogeneous
phase from the charge density wave (CDW) phase, while the one on the
right is the phase boundary in the canonical ensemble (calculated
with $N=84$ and $L=84$). The first order CDW phase transition boundary
is denoted by the double solid line and open squares, while the second
order one is denoted by the solid line and squares. In addition, the
double dashed line denotes the transition boundary for a first order
liquid-gas-like transition without $\mathbb{Z}_{2}$ symmetry breaking
that only exits in the canonical ensemble. Both the black and the
grey dot denote the critical point where the CDW order parameter jump
of the first order transition vanishes. For the canonical ensemble,
the compressibility $\kappa$ of the system is \emph{negative} in
the shaded parameter region. Moreover, the system can reenter the
CDW phase from the homogeneous phase upon \emph{increasing }the temperature,
for instance, along the dashed arrow. (b) Illustration of additivity
breaking in the system of $L$ sites by its ILR interaction. The left
part of the system, with ``even'' sub-lattice sites (gray squares)
homogeneous occupied by one particle (black dot) per site, assumes
the energy $E_{1}=-U_{l}L/16$, while the right part, with ``odd''
sub-lattice sites (white squares) homogeneous occupied by one particle
per site also assumes the energy $E_{2}=-U_{l}L/16$. However, the
total energy of the system $E=0$ is \emph{not} the sum of the energy
of its two parts. (c) Illustration of possible arising of negative
compressibility for generic non-additive systems, whose free energy
$F$ as a function of the total particle number $N$ (solid curve)
can in principle assumes a non-convex part for $N\in(N_{1},N_{2})$
where compressibility $\kappa$ is negative. The dotted line segment
is cotangent to the solid curve at its two ends $(F_{1},N_{1})$ and
$(F_{2},N_{2})$. The upper and lower red dots, with the same total
particle number $\mathcal{N}\in(N_{1},N_{2})$ but different free
energy $\mathcal{F}$ and $\mathcal{F}'$, lie on the solid curve
and the dashed line segment, respectively. See text for more details. }
\label{Fig.canonical_vs_grand_canonical_phase_diagram}
\end{figure}

\begin{figure}
\includegraphics[width=1.76in]{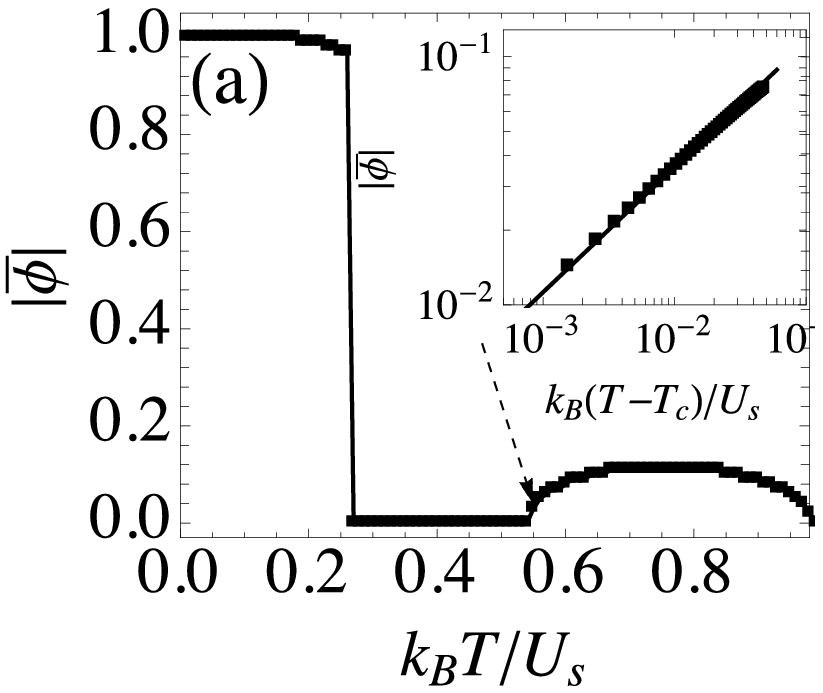}\includegraphics[width=1.6in]{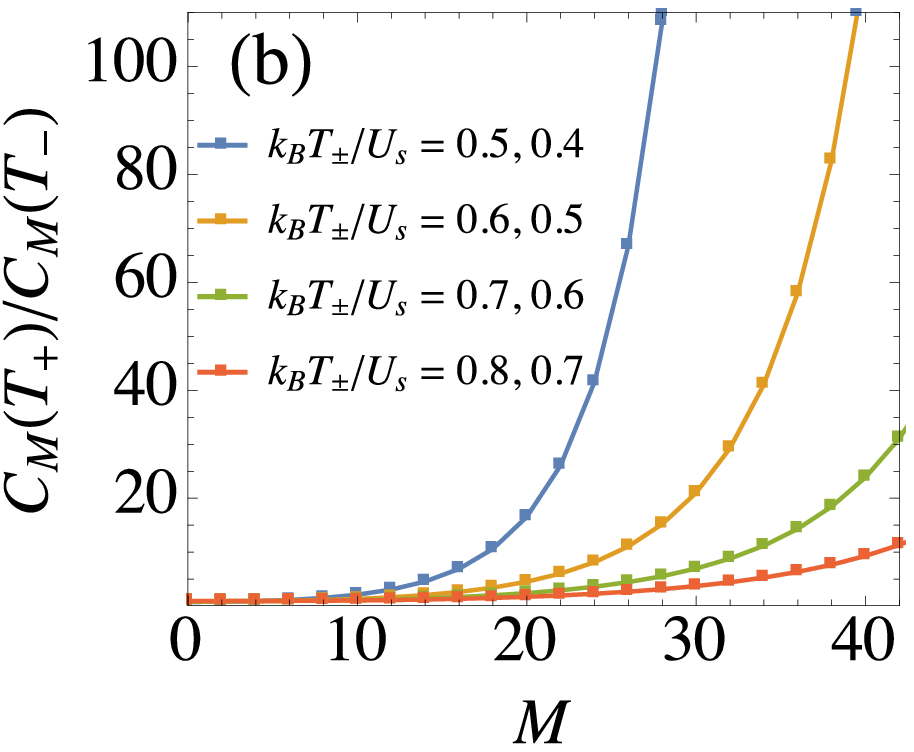}\caption{(a) Temperature dependence of CDW order at fixed $U_{l}=0.65U_{s}$
showing anomalous reentrant second order CDW transition at the temperature
$T_{c}\simeq0.541U_{s}/k_{B}$. Inset: Scaling behavior of $|\bar{\phi}|$
near $T_{c}$, where a scaling law $|\bar{\phi}|\propto(T-T_{c})^{0.498}$
is identified. (b) From left to right, the four curves correspond
to the $M$ dependence of the ratio $C_{M}(T_{+})/C_{M}(T_{-})$ for
four sets of $T_{\pm}$ with $(T_{+},T_{-})=(0.5,0.4),\,(0.6,0.5),\,(0.7,0.6),\,(0.8,0.7)$,
respectively. Obvious monotonically increasing behavior can be identified
for all these four curves, indicating the influence of fluctuations
with larger $|M|$ increases faster with respect to $T$. See text
for more details.}
\label{Fig.Reentrance}
\end{figure}

\begin{figure}
\includegraphics[width=2.7in]{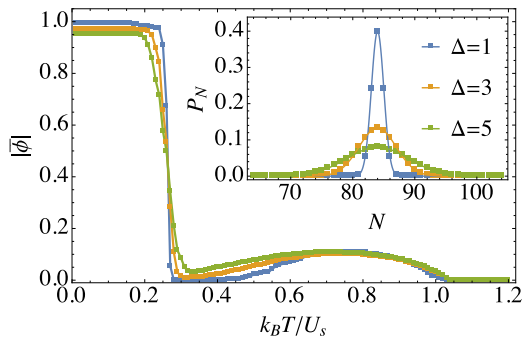}

\caption{``Snapshot'' averaged temperature dependence of $|\bar{\phi}|$
at $U_{l}=0.65U_{s}$. The blue, orange, and blue curves correspond
to $\Delta=1,3,5$ in the ``discrete'' normal distributions $P_{N}$,
respectively. The reentrant transition is still observable in the
when the total particle number fluctuation are relatively small ($\Delta=1,3$
in this case). Inset: Discrete normal distributions $P_{N}$ with
$\Delta=1,3,5$ and $N_{0}=L=84$. See text for more details.}
\label{Fig.Experimental_observability}
\end{figure}

In this work, we address this question for the lattice Bose gases
in optical cavities in the deep Mott-insulator limit at unit filling.
To this end, we establish the finite temperature phase diagram of
the system in the canonical ensemble and find it assumes an anomalous
distinct structure compared to its grand canonical (GC) counterpart
{[}cf.~Fig.~\ref{Fig.canonical_vs_grand_canonical_phase_diagram}(a){]},
which can be traced back to the breaking of system's additivity on
the fundamental level by its ILR interaction {[}cf.~Fig.~\ref{Fig.canonical_vs_grand_canonical_phase_diagram}(b){]}.
More specifically, we find the following: (i) Arising of anomalous
negative compressibility in the canonical ensemble {[}cf.~shaded
region in Fig.~\ref{Fig.canonical_vs_grand_canonical_phase_diagram}(a){]}
and ensemble inequivalence. One direct consequence of additivity breaking
is the possible emergence of negative compressibility as illustrated
in Fig.~\ref{Fig.canonical_vs_grand_canonical_phase_diagram}(c).
Indeed, we find the system manifests negative compressibility in the
canonical ensemble in the temperature regime around half of the on-site
energy. Due to the fact that compressibility is positive semi-definite
in  canonical ensemble irrespective of the absence of additivity,
this thus directly results in the ensemble inequivalence between the
canonical and the GC one, sharply manifesting in the anomalous structure
of the canonical phase diagram as shown in Fig.~\ref{Fig.canonical_vs_grand_canonical_phase_diagram}(a).
(ii) An anomalous reentrant phase transition where upon increasing
the temperature, the ordered charge density wave (CDW) phase revives
from the disordered homogenous phase {[}cf.~Fig.~\ref{Fig.Reentrance}(a)
and the dashed arrow in Fig.~\ref{Fig.canonical_vs_grand_canonical_phase_diagram}(a){]}.
This transition is driven by the cooperation between the ILR interaction
and the particle number fluctuations favored by the CDW phase that
become more influential upon increasing the temperature {[}cf.~Fig.~\ref{Fig.Reentrance}(b){]}.
Moreover, it is robust against small total particle number fluctuations
(cf.~Fig.~\ref{Fig.Experimental_observability}), indicating it
can be readily observed in current experimental set-ups either with
well-controlled total particle number fluctuations, or by post-selection
of experimental data based on the total particle number in each run.

\emph{Model and breaking of additivity.}---In order to describe lattice
Bose gases in optical cavities, we employ the ILR interacting Bose-Hubbard
model which is able to capture their physics in a wide range of the
parameter space \citep{Landig_Nature_2016}. The Hamiltonian of this
model consists of a conventional hopping part and an interaction part.
Here, we concentrate on the physics when the hopping amplitude is
negligibly small, i.e., in the deep Mott-insulator limit, where the
physics of the system is dominated by the interaction part completely
and captured by the Hamiltonian 
\begin{equation}
\hat{H}=\frac{U_{s}}{2}\sum_{i,\sigma}\hat{n}_{i,\sigma}(\hat{n}_{i,\sigma}-1)-\frac{U_{l}}{L}\left(\sum_{i}\hat{n}_{i,e}-\sum_{i}\hat{n}_{i,o}\right)^{2}.\label{eq:Hamiltonian}
\end{equation}
The first term in Eq.~(\ref{eq:Hamiltonian}) is the conventional
short-range on-site interaction term with $U_{s}$ being its strength
and the second term is the cavity-photon-mediated ILR interaction
term \citep{Landig_Nature_2016,Mottl_Science_2012} with its strength
being $U_{l}$. In order to make the total energy of the system still
extensive, hence restore the conventional thermodynamic limit, the
ILR term is further rescaled by the total number of lattice sites
$L$ according to the Kac prescription \citep{Kac_J_Math_Phys_1963}.
To be concrete, we assume the underlying lattice being a 2D square
lattice which is most relevant for the current experimental set-up
\citep{Landig_Nature_2016}. Its two interpenetrating square sub-lattices
are referred as ``even'' ($e$) and ``odd'' ($o$) lattice, respectively.
The number of atoms at site $i$ on the sub-lattice $\sigma$, with
$\sigma$ being $e$ or $o$, are counted by the bosonic particle
number operator $\hat{n}_{i,\sigma}$. On the fundamental symmetry
level the system assumes a $\mathbb{Z}_{2}$-symmetry between these
two sub-lattices \citep{Mottl_Science_2012}. Its associated phase
transition has been observed experimentally in the low temperature
regime \citep{Landig_Nature_2016}, which is a CDW transition between
the homogeneous phase with uniform particle density distribution favored
by the on-site interaction and the CDW phase with particle densities
on the two sub-lattices being different favored by the ILR interaction.

We obtain most of our results by directly calculating system's canonical
partition function $Z_{\mathrm{C}}\equiv\mathrm{tr}[\exp(-(k_{B}T)^{-1}\hat{H})]$
with $k_{B}$ and $T$ being the Boltzmann constant and the temperature,
respectively. The calculation is facilitated by reformulating $Z_{\mathrm{C}}$
as an integral with respect to the CDW order parameter field $\phi$
introduced via the standard Hubbard-Stratonovich transformation \citep{Supplemental_Material}.
This integral form explicitly reads 
\begin{equation}
Z_{\mathrm{C}}=\sqrt{\frac{U_{l}L}{\pi k_{B}T}}\int_{-\infty}^{+\infty}d\phi\,e^{-(k_{B}T)^{-1}Lf_{\{T,U_{s},U_{l}\}}(\phi)}\label{eq:Z_c}
\end{equation}
with 
\begin{align}
 & f_{\{T,U_{s},U_{l}\}}(\phi)\equiv U_{l}\phi^{2}-\frac{k_{B}T}{L}\ln\left(\sum_{M=-N}^{N}e^{\frac{2U_{l}\phi M}{k_{B}T}}C_{M}(T)\right),\label{eq:f_function}
\end{align}
where $C_{M}(T)\equiv\zeta(T)^{-1}\left[\prod_{\sigma}\sum'_{\{n_{i,\sigma}\}}e^{-\frac{U_{s}}{2k_{B}T}\left(\sum_{i}n_{i,\sigma}^{2}-N_{\sigma}\right)}\right]$
and $\zeta(T)\equiv\sum_{M=-N}^{N}\zeta(T)C_{M}(T)$. Here, $n_{i,\sigma}$
is the occupation number, i.e., the eigenvalue of the bosonic particle
number operator $\hat{n}_{i,\sigma}$, and $\sum_{\{n_{i,\sigma}\}}'$
is the summation over all possible value of $\{n_{i,\sigma}\}$ under
the particle number constraints $\sum_{i}n_{i,\sigma}=N_{\sigma}$
and $N_{e}+N_{o}=N$, with $N_{\sigma}$ and $N$ being the total
number of particles on the sub-lattice $\sigma$ and the whole lattice,
respectively. $M$ is the total particle number difference between
the even and odd sub-lattice, i.e., $M\equiv N_{e}-N_{o}$, hence
assumes the value $-N,\,-N+2,\,\cdots,N-2,\,N$. The CDW order parameter
directly equals to the expectation value of $\phi$, denoted as $\bar{\phi}$
in the following, i.e., $\bar{\phi}\equiv\langle\phi\rangle=\langle\sum_{i=1}^{L/2}\hat{n}_{i,e}-\sum_{i=1}^{L/2}\hat{n}_{i,o}\rangle/L$.
In the thermodynamic limit, $Z_{\mathrm{C}}$ can be evaluated exactly
by the saddle point integration and $\bar{\phi}$ is determined by
the value of $\phi$ that minimizes $f_{\{T,U_{s},U_{l}\}}(\phi)$
(see Supplemental Material \citep{Supplemental_Material} for technical
details). We further remark since Hamiltonian (\ref{eq:Hamiltonian})
is completely general for any bipartite lattice, the results presented
below apply generally in all bipartite lattice cases.

Before discussing of our results in more detail, let us now check
yet another fundamental aspect of the system besides its symmetry,
namely, the range of its interaction. In fact, one immediately finds
that the range of its interaction is so long that it directly breaks
its fundamental additivity as illustrated below {[}cf.~Fig.~\ref{Fig.canonical_vs_grand_canonical_phase_diagram}(b){]}.
Let us assume there is a system of size $L$ divided into the left
and the right subsystem with the same size $L/2$, and $N=L/4$ particles
are homogeneously distributed over the even (odd) sub-lattice of the
left (right) subsystem. One can show that the energy of each sub-system
is $-U_{l}L/16$, however, the energy of the whole system is $0$,
clearly not the sum of the sub-system's energy. This thus strongly
suggests that there must exist new physics beyond the scope of the
conventional statistical mechanics where additivity is taken for granted.
Indeed, as we shall see in the following, the breaking of additivity
in the system gives rise to a series of anomalous thermodynamic phenomena,
including negative compressibility, ensemble inequivalence, and anomalous
reentrant phase transitions.

\emph{Negative compressibility and ensemble inequivalence.}---One
of the most direct consequences due to additivity breaking is the
possible arising of negative compressibility. This can be shown generally
by checking the convexity of system's free energy $F$ as a function
of its total particle number $N$ as we now discuss. In the canonical
ensemble, the stable equilibrium state at given total particle number
is the macroscopic state that minimizes the free energy. If the non-convex
$F(N)$ curve in Fig.~\ref{Fig.canonical_vs_grand_canonical_phase_diagram}(c)
described a system with additivity, this system is in fact unstable
with its free energy $\mathcal{F}$ determined by the non-convex part
of $F(N)$ for any total particle number $\mathcal{N}\in(N_{1},N_{2})$.
As illustrated in Fig.~\ref{Fig.canonical_vs_grand_canonical_phase_diagram}(c),
this is because, precisely due to additivity, the system can phase
separate into two subsystems with $(F_{1},N_{1})$ and $(F_{2},N_{2})$
in such a way that it keeps the total particle number $\mathcal{N}$
unchanged, i.e., $\mathcal{N}=\alpha N_{1}+(1-\alpha)N_{2}$ with
$\alpha\in(0,1)$, while lowering its free energy to $\mathcal{F}'=\alpha F_{1}+(1-\alpha)F_{2}$
at the same time by exploiting the non-convex feature of $F(N)$.
However, for a non-additive system, its free energy is generally not
equal to the weighted sum of its subsystem's free energy. Therefore,
in principle, the non-convex $F(N)$ curve shown in Fig.~\ref{Fig.canonical_vs_grand_canonical_phase_diagram}(c)
can describe a stable physical system where no phase separation occurs.
Noticing the compressibility $\kappa\propto(\partial^{2}F/\partial N^{2})^{-1}$
according to Maxwell relations, this thus directly indicates the possible
arising of negative compressibility in non-additive systems, for instance,
those with their additivity broken by their long-range interactions.
Indeed, as shown in the gray region of Fig.~\ref{Fig.canonical_vs_grand_canonical_phase_diagram}(a),
by calculating system's free energy around unit filling (see Supplemental
Material \citep{Supplemental_Material} for technical details), we
find the system assumes negative compressibility in the parameter
regime where $k_{B}T$ and $U_{l}$ assumes similar energy scale as
$U_{s}$.

As a matter of fact, the arising of negative compressibility directly
indicates the ensemble inequivalence between the canonical ensemble
and the GC one, since in the GC ensemble, compressibility is always
positive semi-definite irrespective of additivity breaking \citep{footnote_0}.
This ensemble inequivalence sharply manifests in various anomalous
structures unique to the canonical phase diagram of the system as
shown in Fig.~\ref{Fig.canonical_vs_grand_canonical_phase_diagram}(a).
Besides the negative compressibility regime as discussed previously
and the obvious quantitative difference that for the canonical ensemble,
the CDW phase forms at a noticeably larger $U_{l}$ in the low temperature
regime, the ensemble inequivalence manifests most remarkably in the
temperature regime around half the on-site energy: For the GC ensemble,
the first order CDW transition terminates at a critical point and
changes to a second order one {[}cf.~the phase boundary on the left
(reproduced according to Ref.~\citep{He_arXiv_2020}) in Fig.~\ref{Fig.canonical_vs_grand_canonical_phase_diagram}(a){]}.
While in contrast, for the canonical ensemble, though the first order
CDW transition indeed terminates at the intersection point with the
second order one (more precisely, its anomalous reentrant transition
part, to be discussed later), this intersection point assumes a finite
order parameter jump, hence not a critical point as in the GC case.
In fact, above this intersection point, the first order CDW transition
changes to a first order transition \emph{without} $\mathbb{Z}_{2}$
symmetry breaking, similar to the conventional liquid-gas transition.
This liquid-gas-like transition is characterized by the finite CDW
order parameter jump when the transition boundary is crossed, and
terminates at a critical point at a higher temperature {[}cf.~the
gray dot in Fig.~\ref{Fig.canonical_vs_grand_canonical_phase_diagram}(a){]},
where the jump vanishes. In fact, the second order CDW transition
boundary shows an even more anomalous structure: it assumes an anomalous
part in the temperature regime around $U_{s}/(2k_{B})$, where the
critical $U_{l}$ that drives the CDW transition \emph{decreases}
with respect to $T$. As a direct consequence, this indicates an anomalous
reentrant phase transition, where upon increasing the temperature,
the ordered CDW phase revives from the already disordered homogenous
phase {[}cf. the dashed arrow in Fig.~\ref{Fig.canonical_vs_grand_canonical_phase_diagram}(a){]},
as we shall discuss below in more detail.

\emph{Reentrant phase transition.}---The reentrant transition is
most easily seen by monitoring the temperature dependance of CDW order
as shown in a representative case with $U_{l}=0.65U_{s}$ in Fig.~\ref{Fig.Reentrance}(a).
We observe that upon increasing the temperature, $|\bar{\phi}|$ first
jumps to zero via the first order CDW transition, however revives
from zero to finite value at the critical temperature $T_{c}$ of
the reentrant second order transition. The physical mechanism that
drives this reentrant transition can be attributed to the cooperation
between the long range interaction and the fluctuations with finite
sub-lattice particle number differences $M$ that become more influential
upon increasing $T$ as we now discuss. In the critical regime near
$T_{c}$ of the reentrant transition, the physics of the system is
determined by the properties of $f_{\{T,U_{s},U_{l}\}}(\phi)$ in
the regime $|\phi|\ll1$. Therefore, we can simply focus on the leading
order expansion of $f$ with respect to $\phi$, i.e., $f/U_{l}\simeq\left(1-4U_{l}(k_{B}TL)^{-1}\sum_{M=2}^{N}C_{M}(T)M^{2}\right)\phi^{2}+C$,
with $C$ being an irrelevant constant independent of $\phi$. The
finite CDW order emerges when the coefficient of $\phi^{2}$ in this
expansion becomes negative. This thus indicates if the weighted sum
$\sum_{M=2}^{N}C_{M}(T)M^{2}$ that couples to the long range interaction
$U_{l}$ is a strictly increasing function of $T$, the system can
hold the possibility to reenter the ordered phase upon increasing
$T$. In fact, noticing $C_{M}(T)$ assumes the physical meaning of
the total statistical weight of all the fluctuating configurations
with sub-lattice particle number difference $M$ in the absence of
$U_{l}$, one would expect that although the fluctuations with larger
$|M|$ are suppressed more strongly at low temperature, their influence
will increase faster upon increasing $T$, i.e., one would expect
$\left.C_{M}(T_{+})/C_{M}(T_{-})\right|_{M=M_{1}}>\left.C_{M}(T_{+})/C_{M}(T_{-})\right|_{M=M_{2}}$
for $\forall\,|M_{1}|>|M_{2}|$ with $T_{+}>T_{-}$. If this is the
case, noticing additionally the normalization condition $\sum_{M}C_{M}(T)=1$,
one can show that the weighted sum $\sum_{M=2}^{N}C_{M}(T)M^{2}$
is precisely a strictly increasing function of $T$. Indeed, by directly
calculating $C_{M}(T)$ at different temperature, we find for any
fixed $T_{+}>T_{-}$, $C_{M}(T_{+})/C_{M}(T_{-})$ is a monotonically
increasing function with respect to $|M|$ as shown in a few representative
cases in Fig.~\ref{Fig.Reentrance}(b).

Moreover, the reentrant transition assumes a critical power law scaling,
which we find to be $\bar{\phi}\propto(T-T_{c})^{0.498}$ as shown
in the inset of Fig.~\ref{Fig.Reentrance}(a). On the one hand, this
scaling exponent is numerically the same as the one in the GC ensemble
\citep{He_arXiv_2020}, which belongs to the five dimensional Ising
universality class, clearly bearing the long-range characteristic
of system's interaction. However, on the other hand, in sharp contrast
to the conventional second order CDW transition where $\bar{\phi}$
decreases with respect to temperature in the critical regime, here,
$\bar{\phi}$ \emph{increases} with respect to the temperature in
this critical regime clearly bearing the distinct feature of the reentrant
transition.

\emph{Experimental observability.}---As we have discussed above,
various anomalous behavior appear in the canonical ensemble, i.e.,
under the condition with fixed total particle number, which seems
to be unrealistic for current experiments at first sight. However,
as we shall show now, direct experimental observations are in fact
quite promising, if the total particle number fluctuations of different
runs in experiments can be controlled below a finite small level.
To proceed on a concrete basis, we assume the probability for an experimental
run with total particle number $N$ denoted as $P_{N}$ satisfies
a ``discrete'' normal distribution, i.e., $P_{N}\propto\exp(-(N-N_{0})^{2}/2\Delta^{2})$.
Here, $N_{0}$ is the target total particle number that is supposed
to be achieved in each run of experiments, and $\Delta$ is the variance.
We then directly calculate the properties of the system at different
total particle number and average the results according to the probability
distribution $P_{N}$. Fig.~\ref{Fig.Experimental_observability}
shows three ``snapshot'' averaged temperature dependence of the
CDW order parameter at $U_{l}=0.65U_{s}$ with respect to $P_{N}$
whose $\Delta=1,3,5$, respectively. We can observe that the reentrant
transition still manifests itself in the ``snapshot'' averaged temperature
dependence of $|\bar{\phi}|$ when $\Delta=1,3$, indicating it still
can be observed in experiments when the total particle number fluctuations
are relatively small. Moreover, experimental data with total particle
number lying in the desirable range can generally be post-selected
according to the outcomes of the total particle number measurements.

\emph{Conclusion.}---The breaking of  additivity drastically influences
the fundamental properties of lattice Bose gases in optical cavities:
it directly indicates the arising of negative compressibility in the
canonical ensemble that inevitably breaks ensemble equivalence between
the canonical and GC one. This sharply manifests in the anomalous
structure of the canonical phase diagram, where particularly, in addition
to the region with negative compressibility, it assumes reentrant
CDW phase transitions via which the ordered phases revive from the
disordered ones by heating. We believe our work will stimulate both
further experimental and theoretical investigations on the anomalous
thermodynamic associated to additivity breaking in ultracold atom
systems with long-range interactions.
\begin{acknowledgments}
This work was supported by NSFC (Grant No.~11874017, No.~11674334,
and No.~11947302), GDSTC under Grant No.~2018A030313853, and START
grant of South China Normal University.
\end{acknowledgments}


\begin{thebibliography}{10}
\bibitem{footnote_0} According to Maxwell relations, one can show
$\kappa\propto\partial\langle\hat{N}\rangle_{\mathrm{GC}}/\partial\mu=\beta\left(\langle\hat{N}^{2}\rangle_{\mathrm{GC}}-\langle\hat{N}\rangle_{\mathrm{GC}}^{2}\right)\geq0,$
where $\hat{N}\equiv\sum_{i,\sigma}\hat{n}_{i,o}$, and $\langle\hat{N}\rangle_{\mathrm{GC}}$
($\langle\hat{N}^{2}\rangle_{\mathrm{GC}}$) is the expectation value
for $\hat{N}$($\hat{N}^{2}$) in the GC ensemble.
\end{thebibliography}
\end{document}


\title{Supplemental Material for ``Anomalous thermodynamics of lattice Bose
gases in optical cavities''}
\author{Liang He}
\affiliation{Guangdong Provincial Key Laboratory of Quantum Engineering and Quantum
Materials, SPTE, South China Normal University, Guangzhou 510006,
China}
\author{Su Yi}
\affiliation{CAS Key Laboratory of Theoretical Physics, Institute of Theoretical
Physics, Chinese Academy of Sciences, Beijing 100190, China}
\affiliation{School of Physical Sciences \& CAS Center for Excellence in Topological
Quantum Computation, University of Chinese Academy of Sciences, Beijing
100049, China}
\maketitle

\section{Hubbard-Stratonovich transformation on the canonical partition function}

In the occupation number representation, the explicit form of the
quantum partition function of the system in the canonical ensemble
reads 
\begin{align}
Z_{\mathrm{C}} & =\sum_{\{n_{i,\sigma}\}}e^{-\beta\left\{ \sum_{i,\sigma}\left[\frac{U_{s}}{2}n_{i,\sigma}(n_{i,\sigma}-1)\right]-\frac{U_{l}}{L}\left[\sum_{i}\left(n_{i,e}-n_{i,o}\right)\right]^{2}\right\} },\label{eq:Partition_function_explicit_form}
\end{align}
with $\beta\equiv(k_{B}T)^{-1}$ being the inverse temperature. The
Hubbard-Stratonovich transformation that we use to introduce the CDW
order parameter field $\phi$ into the partition function reads 
\begin{align}
 & \left(\sqrt{\frac{\beta U_{l}L}{\pi}}\right)^{-1}\exp\left(\beta\frac{U_{l}}{L}\left[\sum_{i=1}^{L/2}\left(n_{i,e}-n_{i,o}\right)\right]^{2}\right)\nonumber \\
 & =\int_{-\infty}^{+\infty}d\phi\,\exp\left(-\beta U_{l}\left\{ L\phi^{2}+2\left[\sum_{i=1}^{L/2}\left(n_{i,e}-n_{i,o}\right)\right]\phi\right\} \right).\label{eq:HST}
\end{align}
By using Eq.~(\ref{eq:HST}) we can replace the long range interaction
term appearing in Eq.~(\ref{eq:Partition_function_explicit_form})
by the integral over the $\phi$ field and rewrite the partition function
$Z_{\mathrm{C}}$ in terms of $\phi$ as 
\begin{equation}
Z_{\mathrm{C}}=\sqrt{\frac{U_{l}L}{\pi k_{B}T}}\int_{-\infty}^{+\infty}d\phi\,e^{-(k_{B}T)^{-1}Lf_{\{T,U_{s},U_{l}\}}(\phi)}\label{eq:Z_c_in_sup_mat}
\end{equation}
with 
\begin{align}
 & f_{\{T,U_{s},U_{l}\}}(\phi)\equiv U_{l}\phi^{2}-\frac{k_{B}T}{L}\ln\left(\sum_{M=-N}^{N}e^{\frac{2U_{l}\phi M}{k_{B}T}}C_{M}(T)\right),\label{eq:f_function_in_fup_mat}
\end{align}
as also shown in Eqs.~(2, 3)\textcolor{blue}{{} }in the main text.
The summation involved in the $f_{\{T,U_{s},U_{l}\}}(\phi)$ can not
be performed analytically, however one can calculated numerically
for sufficiently large system. Here, $\phi$ can be regarded as the
fluctuating CDW order parameter filed. Its expectation value $\bar{\phi}\equiv\langle\phi\rangle$
equals to CDW order parameter $\langle\sum_{i=1}^{L/2}\hat{n}_{i,e}-\sum_{i=1}^{L/2}\hat{n}_{i,o}\rangle/L$
exactly, which can be shown by introducing a source $J$ that couples
to $L^{-1}\sum_{i=1}^{L/2}(n_{i,e}-n_{i,o})$ in the partition function
and calculating the derivative $\left.\partial\ln Z_{\mathrm{C}}[J]/\partial J\right|_{J=0}$.

In the thermodynamic limit $L\rightarrow\infty$, the integral with
respect to $\phi$ in Eq.~(\ref{eq:Z_c_in_sup_mat}) is given exactly
by its saddle point integration. Therefore, in the thermodynamic limit,
\begin{equation}
Z_{\mathrm{C}}=\left(\sqrt{\frac{U_{l}L}{\pi k_{B}T}}\right)e^{-(k_{B}T)^{-1}L\min\left[f_{\{T,U_{s},U_{l}\}}(\phi)\right]}\label{eq:Z_C_in_THE_limit}
\end{equation}
 and the CDW order parameter $\bar{\phi}$ is given by the value of
$\phi$ that minimizes $f_{\{T,U_{s},U_{l}\}}(\phi)$.

\section{Caculation of Free energy and compressibility}

In order to check the positiveness of the compressibility, we need
to calculate second order derivative of free energy $F$with respect
to the total particle number $N$, i.e., $\partial^{2}F/\partial N^{2}$.
This calculation is facilitated by using the expressioin of canonical
partition function shown in Eq.~(\ref{eq:Z_C_in_THE_limit}), via
which we can get 
\begin{equation}
F=-k_{B}T\ln Z_{\mathrm{C}}=-k_{B}T\ln\sqrt{\frac{\beta U_{l}L}{\pi}}+L\min\left[f_{\{T,U_{s},U_{l}\}}(\phi)\right].
\end{equation}
Therefore, by directly calculating $\min\left[f_{\{T,U_{s},U_{l}\}}(\phi)\right]$
at different total particle number $N$ of the system, we can construct
the function $F(N)$ and calculate its second order derivative $\partial^{2}F/\partial N^{2}$.